\documentstyle[aps,prb,epsf, multicol]{revtex}
\tighten
\begin{document}
\draft
\newcommand{\bn}{{\bf n}}
\newcommand{\bp}{{\bf p}}
\newcommand{\br}{{\bf r}}
\newcommand{\bq}{{\bf q}}
\newcommand{\bj}{{\bf j}}
\newcommand{\bE}{{\bf E}}
\newcommand{\eps}{\varepsilon}
\newcommand{\la}{\langle}
\newcommand{\ra}{\rangle}
\newcommand{\cK}{{\cal K}}
\newcommand{\cD}{{\cal D}}
\newcommand{\mybeginwide}{
    \end{multicols}\widetext
    \vspace*{-0.2truein}\noindent	    
    \hrulefill\hspace*{3.6truein}
}
\newcommand{\myendwide}{
    \hspace*{3.6truein}\noindent\hrulefill
    \begin{multicols}{2}\narrowtext\noindent
}

\title{
  Cascade approach to current fluctuations in a chaotic cavity
}
\author{
   K. E. Nagaev$^{1,2}$, P. Samuelsson$^2$, and S. Pilgram$^2$
}
\address{
  $^1$Institute of Radioengineering and Electronics,
  Russian Academy of Sciences, Mokhovaya ulica 11, 101999 Moscow,
  Russia\\
  $^2$D\'epartement de Physique Th\'eorique, Universit\'e de 
  Gen\'eve, CH-1211, Gen\'eve 4, Switzerland\\
}
  
\maketitle

\bigskip
\begin{abstract}
We propose a simple semiclassical method for calculating higher-order 
cumulants of current in multichannel mesoscopic
conductors. To demonstrate its efficiency, we calculate the third and
fourth cumulants of current for a chaotic cavity  with
multichannel leads of arbitrary transparency and compare the results
with ensemble-averaged quantum-mechanical quantities. We also explain
the discrepancy between the quantum-mechanical results and previous
semiclassical calculations.  
\end{abstract}

\begin{multicols}{2}
\narrowtext
\vspace{1cm}

\section{
                       INTRODUCTION
}

In the last decade, there has been a large interest in current 
correlations in mesoscopic conductors.\cite{Blanter-00a} Recently, also 
higher cumulants of current have received a significant attention of 
theorists. In a series of papers\cite{Levitov-93,Lee-95,Levitov-96} a 
scattering approach to the distribution of charge transmitted through 
an arbitrary multi-terminal, multi-mode mesoscopic conductor, i.e. the 
so-called full counting statistics has been developed. As was shown in 
Ref.\cite{Lee-95}, the ensemble-averaged cumulants of arbitrary order can 
be calculated for any two-terminal conductor where the distribution 
of the transmission eigenvalues is known, e.g. for diffusive
wires\cite{Mello-89}, chaotic cavities,\cite{Jalabert-94,Baranger-94,%
Brouwer-96}, double-barrier tunnel junctions\cite{deJong-96a}, or
combinations of different conductors.\cite{Nazarov-94} 

Recently, Nazarov\cite{Nazarov-99} presented a method for 
calculating the full counting statistics of charge transfer in
conductors with a large number of quantum channels based on equations
for the semiclassical Keldysh Green's functions. Subsequently, this 
method was extended\cite{Nazarov-02} to multiterminal systems. Also,
other approaches to higher cumulants, such as the nonlinear sigma
model\cite{Gutman-02} for diffusive wires, have been proposed. 

Common to all approaches\cite{Levitov-93,Lee-95,Levitov-96,Nazarov-99,%
Nazarov-02,Gutman-02} is that they are based on a quantum 
mechanical formulation. To obtain the cumulants for semiclassical 
systems, i.e. systems much larger than the Fermi wavelength, an
ensemble average is performed and the number of transport modes is 
set to infinity, i.e. single-mode weak-localization-like corrections
are neglected. Therefore it is of interest to have a completely 
semiclassical theory for the higher cumulants, which does not
involve any quantum-mechanical quantities.

A step in this direction was made by de Jong\cite{deJong-96a}, who
calculated the distribution of charge transmitted through a
double-barrier tunnel junction by applying a master equation to the
transport in each completely independent transverse quantum
channel. The results were in agreement with the quantum-mechanical
theory in the limit of large channel number. 

An attempt to construct a fully semiclassical theory of higher
cumulants of current in a chaotic cavity was made by Blanter, 
Schomerus, and Beenakker\cite{Blanter-01}
based on the principle of {\it minimal correlations}.\cite{Blanter-00b}
According to this principle, the fluctuations of the semiclassical
distribution function of electrons in the cavity and the fluctuations
of outgoing currents are related only through the condition of
electron-number conservation, which is equivalent to the 
dephasing-voltage-probe approach\cite{deJong-96} in quantum
mechanics. However an attempt to extend the
minimal-correlation approach to the fourth cumulant has led to a
discrepancy with quantum-mechanical results,\cite{Blanter-00b} which
highly surprised the authors.\cite{Oberholzer-01}

Meanwhile the correlations imposed by the particle-number conservation
are not the only possible ones in semiclassics. Quite recently the
semiclassical Boltzmann--Langevin approach\cite{Kogan-69} has been
extended to higher cumulants.\cite{Nagaev-02} This extension takes
into account the effect of lower cumulants on higher cumulants through
the fluctuations of the distribution function and therefore it was
termed {\it cascade} approach. Its equivalence with quantum-mechanical
results\cite{Lee-95}  has been proven for diffusive metallic
conductors.\cite{Nagaev-02} In this paper, we show that the cascade 
approach is not restricted to diffusive metals or to conductors
where the scattering is described by a collision integral, but it may be
also applied to other systems that allow a semiclassical description, 
e.g. to chaotic cavities. To this end, we semiclassically
calculate the third and fourth cumulants of current in a chaotic
cavity taking into account cascade correlations and show that these
values coincide with ensemble-averaged quantum-mechanical results. 

The paper is organized as follows. In Section \ref{Model} we describe
the model of chaotic cavity to be considered. The minimal-correlation
results for the second cumulant of current are presented in Section
\ref{Minimal}. In Section \ref{Cascade} we
\begin{figure}[t]
\epsfxsize6cm \centerline{\epsffile{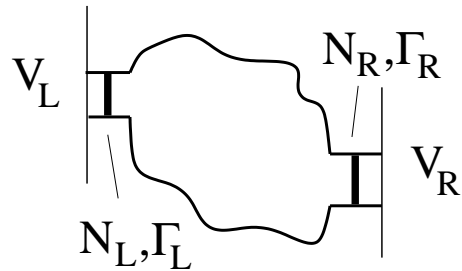}}
\vspace{5mm}
\caption{%
A chaotic cavity with imperfect leads
}
\label{cavity}
\end{figure}
\noindent
calculate the third and
fourth cumulants  of current by means of the cascade approach. In
Section \ref{Quantum}, the third cumulant of current is calculated by
means of the quantum-mechanical circuit theory and its equivalence
with the cascade results is shown. Section \ref{Summary} presents a
conclusion where the results are summarized.

\section{
                           THE MODEL
}\label{Model}
 
Consider a chaotic cavity with two contacts of arbitrary
transparency. The left contact has $N_L \gg 1$ channels and
transparency $\Gamma_L$, and the right contact has $N_R$ channels and
a transparency $\Gamma_R$. The conductances of the leads  $G_L = 
(e^2/\pi\hbar)N_L\Gamma_L$ and $G_R = (e^2/\pi\hbar)N_R\Gamma_R$ are 
also assumed to be much larger than $e^2/\hbar$, and the total 
conductance of the system is that of two resistors connected in 
series\cite{Beenakker-97}
$$
 G
 =
 \frac{
   G_L G_R
 }{
   G_L + G_R
 }.
$$
Because of strong chaotic scattering in the cavity, the electrons
entering the cavity lose the memory of their phase on the time scale of
the order of the time of flight through the cavity yet retain their
energy. Therefore despite the quantum nature of the contacts, the cavity
is a semiclassical object in the sense that the electrons inside it
may be described by a semiclassical distribution function that
depends only on the electron energy. Its average value is given by an
expression\cite{vanLangen-97} 
\begin{equation}
 f(\eps)
 =
 \frac{
   G_L f_L(\eps) + G_R f(\eps)
 }{
   G_L + G_R
 },
 \label{f_average}
\end{equation}
where $f_L(\eps)$ and $f_R(\eps)$ are the distribution functions in the
left and right electrodes.

\section{
           The principle of minimal correlations
}\label{Minimal}

If the distribution function in the cavity were not allowed to
fluctuate, the cavity could be considered just as a reservoir with
non-Fermian distribution of electrons. The contacts would be independent
generators of current noise and the cumulants of the corresponding
extraneous noise currents could be obtained by differentiating the
corresponding quantum-mechanical characteristic
functions\cite{Muzykantskii-94} for the charge transmitted in time $t$

$$
 \chi_{L,R}(\lambda, t)
 =
 \exp
 \biggl\{
   \frac{
      t N_{L,R}
   }{
      2\pi\hbar
   }
   \int d\eps
   \ln
   \bigl\{
     1
$$ $$
     +
     \Gamma_{L,R}
      f(\eps)[1 - f_{L,R}(\eps)]
     ( 
       e^{i\lambda} - 1 
     )
$$ \begin{equation}
     +
     \Gamma_{L,R}
     f_{L,R}(\eps)[1 - f(\eps)]
     (
       e^{-i\lambda} - 1
     )
  \}
 \biggr\}
 \label{chi}
\end{equation}
with respect to the parameter $\lambda$ the corresponding number of
times. 

In what follows, we will be interested only in the Fourier
transforms of the current cumulants in the low-frequency limit and it
will be implied that all the subsequent equations contain only
low-frequency Fourier transforms of the corresponding quantities.
Equation (\ref{chi}) leads to the following expressions for the 
cumulants of the noise current generated by the contacts:
\begin{equation} 
 \la\la
  \tilde I_{L,R}^n
 \ra\ra
 =
 \int d\eps\,
 \la\la
  \tilde I_{L,R}^n
 \ra\ra_{\eps},
 \label{ext-n}
\end{equation}
where
$$
 \la\la
  \tilde I_{L,R}^2
 \ra\ra_{\eps}
 =
 G_{L,R}
 [
  f_{L,R}(1 - f)
  +
  f(1 - f_{L,R})
$$ \begin{equation}
  -
  \Gamma_{L,R}
  (f_{L,R} - f)^2
 ],
 \label{ext-2}
\end{equation}
$$
 \la\la
  \tilde I_{L,R}^3
 \ra\ra_{\eps} 
 =
 eG_{L,R}
 (f - f_{L,R})
 \{
   1
   -
   3\Gamma_{L,R}
   [
    f(1 - f_{L,R})
$$ \begin{equation}
    +
    f_{L,R}(1 - f)
   ]
   +
   2\Gamma_{L,R}^2
   (f - f_L)^2
 \},
 \label{ext-3}
\end{equation}
and
$$
 \la\la
  \tilde I_{L,R}^4
 \ra\ra_{\eps} 
 =
 e^2G_{L,R}
 \Bigl\{
  f_{L,R}(1 - f)
  +
  f(1 - f_{L,R})
$$ $$
  +
  \Gamma_{L,R}
  (
   12f_{L,R}^2 f + 12f^2 f_{L,R} - 12f_{L,R}^2 f^2  
$$ $$
   - 7f_{L,R}^2 -7f^2 + 2ff_{L,R}
  )
$$ $$
  +
  12\Gamma_{L,R}^2
  (f_{L,R} - f)^2
  [
    f_{L,R}(1 - f) + f(1 - f_{L,R})
  ]
$$ \begin{equation}
  -
  6\Gamma_{L,R}^3
  (f_{L,R} - f)^4
 \Bigr\}.
 \label{ext-4}
\end{equation}

Since we are interested here only in low-frequency fluctuations, the
pile-up of electrons in the cavity is forbidden. On the other hand,
the noise currents $\tilde I_L$ and $\tilde I_R$ are absolutely
independent, which would apparently result in a violation of the
current-conservation law if these were the only contributions to the
current noise. To ensure the current conservation at low
frequencies, one has to take into account fluctuations of the
distribution function $\delta f(\eps)$ in the cavity.\cite{vanLangen-97}
Now the fluctuations of the current outgoing from the cavity to the
left and right electrodes assume a form of Langevin equations, where
$\tilde I_L$ and $\tilde I_R$ play the role of extraneous sources.
\begin{equation}
 \delta I_{L,R}
 =
 \tilde I_{L,R}
 +
 \frac{1}{e}
 G_{L,R}
 \int d\eps \delta f(\eps).
 \label{dI}
\end{equation}
Extracting $\delta f$  from the condition of current conservation
$$
 \delta I_L + \delta I_R = 0,
$$
one obtains
\begin{equation}
 \delta I_L
 =
 \frac{
   G_R\tilde I_L - G_L\tilde I_R
 }{
   G_L + G_R
 }.
 \label{dI_L}
\end{equation}
By squaring this equation and using the independence of $\tilde I_L$
and $\tilde I_R$, one easily obtains that the second cumulant of 
the measurable current is
\begin{equation}
 \la\la
  I_L^2
 \ra\ra
 =
 \frac{
    G_R^2
    \la\la \tilde I_L^2 \ra\ra
    +
    G_L^2
    \la\la \tilde I_R^2 \ra\ra
 }{
    ( G_L + G_R )^2
 }.
 \label{I2}
\end{equation}
In the zero-temperature limit it gives
$$
 \la\la
  I_L^2
 \ra\ra
 =
 eI
 \bigl[
  G_L G_R (G_L + G_L)
  +
  G_L^3(1 - \Gamma_R)
$$ \begin{equation}
  +
  G_R^3(1 - \Gamma_L)
 \bigr]
 \left/
  (G_L + G_R)^3
 \right.,
 \label{I2-0}
\end{equation}
where $I$ is the average current flowing through the cavity. In the
high-transparency limit $\Gamma_L = \Gamma_R = 1$ it reproduces the
expression obtained by Blanter and Sukhorukov by means of the
minimal-correlation principle and the exact quantum-mechanical
results. 

\section{
                  Cascade corrections
}\label{Cascade}
  
A straightforward extension of the minimal correlation approach to
higher cumulants has led to a discrepancy with the quantum mechanical
results.\cite{Blanter-01}  The reason
is that the cavity is not just a reservoir with a nonequlibrium
distribution of electrons. As suggested by Eqs. (\ref{dI}), their
distribution function $f(\eps)$ also exhibits fluctuations.  As the
cumulants of the currents $\tilde I_L$ and $\tilde I_R$ are
functionals of the distribution function in the cavity, its
fluctuation $\delta f$ changes them too. Since the characteristic time
scale for $\delta f$ is of the order of the dwell time of an electron
in the cavity, these changes are slow on the scale of the
correlation time of extraneous currents, and therefore the cumulants
of these currents adiabatically follow $\delta f$. This results in
additional correlations, which may be termed ``cascade'' because
lower-order correlators of extraneous currents contribute to
higher-order cumulants of measurable quantities. One can write for the
low-frequency transforms of the corresponding quantities
\begin{equation}
 \delta
  \la\la
  \tilde I_{L,R}^n
 \ra\ra
 =
 \int d\eps\,
 \frac{
   \delta
   \la\la
    \tilde I_{L,R}^n
   \ra\ra
 }{
   \delta f(\eps)
 }
 \delta f(\eps),
 \label{dI^n}
\end{equation}
where $\delta\la\ldots\ra/\delta f$ denotes a functional derivative of
the corresponding quantity with respect to $f(\eps)$. For example, the
third cumulant of the current may be written as the sum of the
minimal-correlation value
\begin{equation}
 \la\la
  I_L^3
 \ra\ra_m
 =
 \frac{
    G_R^3
    \la\la \tilde I_L^3 \ra\ra
    -
    G_L^3
    \la\la \tilde I_R^3 \ra\ra
 }{
    ( G_L + G_R )^3
 }
 \label{I3m}
\end{equation}
and the cascade correction
\begin{figure}[t]
\epsfxsize6cm \centerline{\epsffile{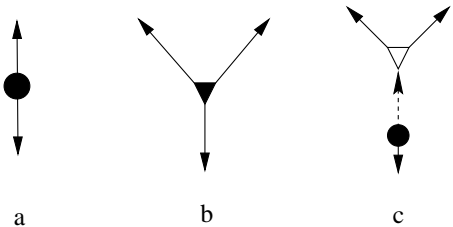}}
\vspace{3mm}
\caption{%
The second cumulant and the two contributions to the third cumulant of
current. The external ends correspond to current fluctuations at
different moments of time and the dashed lines, to fluctuations of the
distribution function. The full circle and triangle correspond to
cumulants of extraneous currents and the empty triangle to the functional
derivative of the second cumulant.
}
\label{cum3}
\end{figure}\noindent
\begin{equation}
 \Delta
 \la\la
  I_L^3
 \ra\ra
 =
 3\int d\eps\,
 \frac{
   \delta  \la\la I_L^2 \ra\ra
 }{
   \delta f(\eps)
 }
 \la
   \delta f(\eps)
   \delta I_L
 \ra.
 \label{DI3}
\end{equation}
The factor 3 is due to the fact that this equation in general includes
three different currents and allows three inequivalent permutations of
them. 
 
The cascade corrections are conveniently presented in a diagrammatic
form\cite{Nagaev-02} (see Figs. \ref{cum3} and \ref{cum4}). The rules
for constructing these diagrams strongly differ from the ones known
for Green's functions in quantum mechanics. The diagrams do not
present an expansion in any small parameter and their number is
strictly limited for a cumulant of a given order. All diagrams present
graphs, whose outer vertices correspond to different instances of
current and whose inner vertices correspond either to cumulants of
extraneous currents or their functional derivatives. The number of
arrows outgoing from an inner vertex corresponds to the order of the
cumulant and the number of incoming arrows corresponds to the order of
a functional derivative. Since the $n$th cumulant presents a
polynomial of the distribution function of degree $n$, the number of
incoming arrows at any inner vertex cannot exceed the number of outgoing
arrows. Apparently, the difference between the total order of
cumulants involved and the total number of functional differentiations
should be equal to the order of the cumulant being calculated. As
there should be no back-action of higher cumulants on lower cumulants,
all diagrams are singly connected. Therefore any diagram for the $n$th
cumulant of the current may be obtained from a diagram of order $m<n$
by combining it with a diagram of order $n-m+1$, i.e. by inserting one
of its outer vertices into one of the inner vertices of the
latter. Hence the most convenient way to draw diagrams for a cumulant
of a given order is to start with diagrams of lower order and to
consider all their inequivalent combinations that give diagrams of
the desired order. The analytical expressions corresponding to each
diagram contain numerical prefactors equal to the numbers of
inequivalent permutations of the outer vertices. 

Unlike the case of a diffusive conductor, the third and fourth
cumulants include now {\it all} possible diagrams and not only those
that are constructed of second-order cumulants (Fig. \ref{cum3},
diagram $a$). The third cumulant is presented by diagrams $b$ and $c$ 
in Fig. \ref{cum3}. Diagram $b$  presents the minimal-correlation
value (\ref{I3m}) and diagram $c$ presents the only possible cascade
correction (\ref{DI3}) obtained by combining two second cumulants. 

We are now in position to evaluate the diagrams.
The functional derivative is easily obtained by differentiating
Eq. (\ref{ext-2}) and substituting it into (\ref{I2}), which gives
$$
 \frac{
   \delta  \la\la I_L^2 \ra\ra
 }{
   \delta f(\eps)
 }
 =
 \frac{
   G_L G_R
 }{
   (G_L + G_R)^2
 }
 \Bigl\{
  G_L
  \bigl[
   1 - 2f_R
   +
   2\Gamma_R
   (f_R - f)
  \bigr]
$$ \begin{equation}
  +
  G_R
  \bigl[
   1 - 2f_L
   +
   2\Gamma_L
   (f_L - f)
  \bigr]
 \Bigr\}.
 \label{dI2/df}
\end{equation}

To calculate the fluctuation $\delta f$, one has to write down
equations (\ref{dI}) and the current-conservation law in the
energy-resolved form. This immediately gives
\begin{equation}
 \delta f(\eps)
 =
 -
 \frac{e}{ G_L + G_R }
 \left[
  (\tilde I_L)_{\eps}
  +
  (\tilde I_R)_{\eps}
 \right]
 \label{df}
\end{equation}
where $(\tilde I_{L,R})_{\eps}$ are energy-resolved extraneous
currents. Since fluctuations at different energies are completely
independent, one easily obtains that
\begin{equation}
 \la
  \delta f(\eps)
  \delta I_L
 \ra
 =
 \frac{e}{ 
     (G_L + G_R)^2
 } 
 \left[
  G_L
  \la\la
   \tilde I_R^2
  \ra\ra_{\eps}
  -
  G_R
  \la\la
   \tilde I_L^2
  \ra\ra_{\eps}
 \right].
 \label{dfdI}
\end{equation}
Hence the total third cumulant, which is the sum of (\ref{I3m}) and
(\ref{DI3}), is of the form
$$
 \la\la
  I_L^3
 \ra\ra
 =
 -
 \frac{
  e^2 I
 }{
  (G_L + G_R)^6
 }
 \Bigl\{
  (G_L + G_R)
  \bigl[
    (G_L + G_R)^2 
$$ $$
    \times
    (G_L^3 + G_R^3)
    -
    3(G_L + G_R)
    (\Gamma_L G_R^4 + \Gamma_R G_L^4)
$$ $$
    +
    2\Gamma_L^2 G_R^5
    +
    2\Gamma_R^2 G_L^5
  \bigr]
  -
  3G_L G_R
  \bigl[
    G_L^2(1 - \Gamma_R)
    -
    G_R^2(1 - \Gamma_L)
  \bigr]
$$ \begin{equation}
  \times
  \bigl[
    G_L^2(1 - 2\Gamma_R)
    -
    G_R^2(1 - 2\Gamma_L)
  \bigr]
 \Bigr\}.
 \label{I3-0}
\end{equation}
In the case of perfectly transparent leads $\Gamma_L =
\Gamma_R = 1$ the cascade correction to the third cumulant is zero and
the minimal-correlation result
$$
 \la\la
  I_L^3
 \ra\ra
 =
 -e^2 I
 \frac{
   G_L G_R
   (G_L - G_R)^2
 }{
   (G_L + G_R)^4
 }
$$
is reproduced. This is why the discrepancy between the
minimal-correlation and quantum-mechanical results was noted by
Blanter and co-workers only for the fourth cumulant.


The fourth cumulant is presented by a sum of six diagrams shown in
Fig. \ref{cum4}. Diagram $a$ presents the minimal-correlation 
value
\begin{figure}[t]
\epsfxsize8cm \centerline{\epsffile{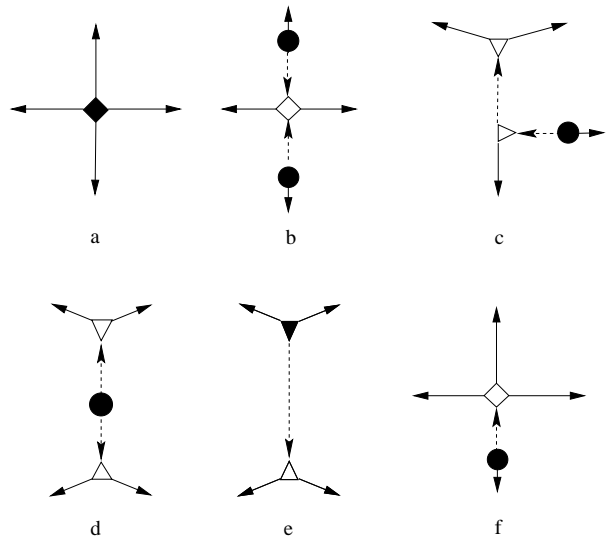}}
\vspace{3mm}
\caption{%
The contributions to the fourth cumulant of the current. Dashed lines
correspond to fluctuations of the distribution function. Full circles,
triangles and squares correspond to the second, third, and fourth
cumulants of extraneous currents. The empty triangles and squares
present their functional derivatives.
}
\label{cum4}
\end{figure}
\begin{equation}
 \la\la
  I_L^4
 \ra\ra_m
 =
 \frac{
    G_R^4
    \la\la \tilde I_L^4 \ra\ra
    +
    G_L^4
    \la\la \tilde I_R^4 \ra\ra
 }{
    ( G_L + G_R )^4
 }.
 \label{I4m}
\end{equation}
The rest of diagrams are obtained by combining the diagrams for the
third and second cumulants. Diagrams $b$ and $c$ are obtained by
inserting the second cumulant into the diagram $c$ for the third 
cumulant, and diagram $d$ is obtained by plugging diagram $c$
for the third cumulant into the second cumulant. Diagrams $e$ and $f$
are obtained by inserting diagram $b$ into diagram $a$ and diagram $a$
into diagram $b$. The corresponding analytical
expressions contain numerical prefactors 1, 6, 12, 3, 6, and 4 that
present the numbers of inequivalent permutations of four currents
entering into the cumulant. 

The first correction is given by an expression
$$
 \Delta_1
 \la\la
  I_L^4
 \ra\ra
 =
 6\int d\eps_1\int d\eps_2\,
 \frac{
   \delta^2
   \la\la I_L^2 \ra\ra
 }{
   \delta f(\eps_1)
   \delta f(\eps_2)
 }
$$ \begin{equation}
 \times
 \la
  \delta f(\eps_1)
  \delta I_L
 \ra
 \la
  \delta f(\eps_2)
  \delta I_L
 \ra.
 \label{D1I4}
\end{equation}
The second functional derivative
\begin{equation}
 \frac{
   \delta^2
   \la\la I_L^2 \ra\ra
 }{
   \delta f(\eps_1)
   \delta f(\eps_2)
 }
 =
 -2\delta(\eps_1 - \eps_2)
 \frac{
   G_L G_R
   (G_R\Gamma_L + G_L\Gamma_R)
 }{
   (G_L + G_R)^2
 }
 \label{d2I2/df2}
\end{equation}
is obtained by differentiating (\ref{I2}) twice with respect to
$f(\eps)$, and the two correlators in (\ref{D1I4}) are given by
(\ref{dfdI}). 

The second cascade correction is given by an expression
$$
 \Delta_2
 \la\la
  I_L^4
 \ra\ra
 =
 12\int d\eps_1\,
 \frac{
   \delta
   \la\la I_L^2 \ra\ra
 }{
   \delta f(\eps_1)
 }
 \int d\eps_2\,
 \frac{
   \delta
   \la \delta f(\eps_1) \delta I_L \ra
 }{
   \delta f(\eps_2)
 }
$$ \begin{equation}
 \times
 \la
  \delta f(\eps_2)
  \delta I_L
 \ra,
 \label{D2I4}
\end{equation}
where the first functional derivative is given by (\ref{dI2/df}), 
$$
 \frac{
   \delta
   \la \delta f(\eps_1) \delta I_L \ra
 }{
   \delta f(\eps_2)
 }
 =
 2\delta(\eps_1 - \eps_2)
 \frac{
   G_L G_R
 }{
   (G_L + G_R)^2
 }
 \bigl[
  (\Gamma_L - \Gamma_R)f
$$ \begin{equation}
  + 
  (1 - \Gamma_L)f_L
  -
  (1 - \Gamma_R)f_R
 \bigr],
 \label{ddfdI/df}
\end{equation}
and the last correlator is given by (\ref{dfdI}). 

The third contribution is given by an expression
\begin{equation}
 \Delta_3
 \la\la
  I_L^4
 \ra\ra
 =
 3\int d\eps_1\int d\eps_2\,
 \frac{
   \delta
   \la\la I_L^2 \ra\ra
 }{
   \delta f(\eps_1)
 }
 \la
   \delta f(\eps_1)
   \delta f(\eps_2)
 \ra
 \frac{
   \delta
   \la\la I_L^2 \ra\ra
 }{
   \delta f(\eps_2)
 },
 \label{D3I4}
\end{equation}
where the functional derivatives are given by (\ref{dI2/df}) and the
second cumulant of the distribution function
\begin{equation}
 \la
  \delta f(\eps_1)
  \delta f(\eps_2)
 \ra
 =
 e^2
 \delta(\eps_1 - \eps_2)
 \frac{
   \la\la
    \tilde I_L^2
   \ra\ra_{\eps_1}
   +
   \la\la
    \tilde I_R^2
   \ra\ra_{\eps_1}
 }{
   (G_L + G_R)^2
 }
 \label{dfdf}
\end{equation}
is obtained by multiplying equations (\ref{df}) with $\eps = \eps_1$
and $\eps = \eps_2$.

The fourth cascade correction involves third-order cumulants of
extraneous currents and is given by an expression
\begin{equation}
 \Delta_4
 \la\la
  I_L^4
 \ra\ra
 =
 6\int d\eps\,
 \frac{
   \delta
   \la\la I_L^2 \ra\ra
 }{
   \delta f(\eps)
 }
 \la
   \delta f(\eps)
   \delta I_L^2
 \ra_m,
 \label{D4I4}
\end{equation}
where
\begin{equation}
 \la
   \delta f(\eps)
   \delta I_L^2
 \ra_m
 =
 -e
 \frac{
   G_R^2
   \la\la
    \tilde I_L^3
   \ra\ra_{\eps}
   +
   G_L^2
   \la\la
    \tilde I_R^3
   \ra\ra_{\eps}
 }{
   (G_L + G_R)^3
 }
 \label{dfdI2}
\end{equation}
is obtained by multiplying one equation (\ref{df}) and two equations
(\ref{dI}) and averaging them with the correlators (\ref{ext-3}).

The fifth correction is given by
\begin{equation}
 \Delta_5
 \la\la
  I_L^4
 \ra\ra
 =
 4\int d\eps\,
 \frac{
   \delta
   \la\la I_L^3 \ra\ra_m
 }{
   \delta f(\eps)
 }
 \la
   \delta f(\eps)
   \delta I_L
 \ra.
 \label{D5I4}
\end{equation}
The functional derivative in the integrand
$$
 \frac{
   \delta
   \la\la I_L^3 \ra\ra_m
 }{
   \delta f(\eps)
 }
 =
 e
 \frac{
  G_L G_R
 }{
  (G_L + G_R)^3
 }
 \Bigl\{
  G_R^2
  \bigl[
   1
   -
   6\Gamma_L f(1-f)
$$ $$
   -
   6\Gamma_L
   (1 - \Gamma_L)
   (f - f_L)^2
  \bigr]
$$ \begin{equation}
  -
  G_L^2
  \bigl[
   1
   -
   6\Gamma_R f(1-f)
   -
   6\Gamma_R
   (1 - \Gamma_R)
   (f - f_L)^2
  \bigr]
 \Bigr\}
 \label{dI3/df}
\end{equation}
is calculated similarly to (\ref{dI2/df}).

The total fourth cumulant of current is given by the sum of its
minimal-correlation value (\ref{I4m}) and the cascade 
\begin{figure}[t]
\epsfxsize5cm \centerline{\epsffile{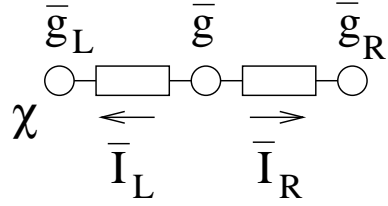}}
\vspace{5mm}
\caption{%
The circuit theory representation of the junction, with 
matrix Greens functions $\bar g_L, \bar g_R$ and $\bar g$, matrix 
currents $\bar I_R$ and $\bar I_L$ and the counting field $\chi$ shown.
}
\label{circuit}
\end{figure}\noindent
corrections
given by (\ref{D1I4}), (\ref{D2I4}), (\ref{D3I4}), (\ref{D4I4}), and
(\ref{D5I4}). The full resulting expression of rather complicated form
is given in the Appendix, and here we give only its limiting values
$$
 \la\la
  I_L^4
 \ra\ra
 =
 e^3 V
 \frac{ 
   G_L^2 G_R^2
 }{
   (G_L + G_R)^7
 }
 \bigl[
  G_L^4 
  - 
  8 G_R G_L^3 
  + 
  12 G_L^2 G_R^2
$$ \begin{equation}
  -
  8 G_L G_R^3
  +
  G_R^4
 \bigr]
 \label{I4-1}
\end{equation}
in the high-transparency limit $\Gamma_L = \Gamma_R = 1$ and
$$
 \la\la
  I_L^4
 \ra\ra
 =
 e^3 V
 \frac{ 
   G_L G_R
 }{
   (G_L + G_R)^7
 }
 \bigl[
  G_L^6
  -
  8 G_L^5 G_R
$$ \begin{equation}
  +
  31 G_L^4 G_R^2
  -
  40 G_L^3 G_R^3
  +
  31 G_L^2 G_R^4
  -
  8 G_L G_R^5
  +
  G_R^6
 \bigr]
 \label{I4-0}
\end{equation}
in the low-transparency limit where $\Gamma_L \to 0$ and $\Gamma_R \to
0$. 

\section{
                Circuit-theory results
}\label{Quantum}

We now show that the same results for the cumulants can be obtained
within an ensemble-averaged quantum mechanical Green's function
approach. We use the circuit theory of full counting statistics,
recently developed by Nazarov and Bagrets, \cite{Nazarov-02} which 
allows us to calculate cumulant by cumulant in a systematic way. For
shortness of the presentation, we only calculate the first three 
cumulants.

The circuit theory representation of the junction is shown in Fig
\ref{circuit}. It consists of three ``nodes'', the two reservoirs and
the dot, connected via two ``resistances'', the left and right point
contact. Each node is assigned a $2\times 2$ matrix Green's function,
i.e. $\bar g_L, \bar g_R$ and $\bar g$. The matrices are subjected to
a normalization condition $\bar g_L^2=\bar g_R^2=\bar g^2=1$. The
Green's functions of the two reservoirs are known,
\begin{eqnarray}
\bar g_{L}&=&e^{i\chi\bar \tau_z/2} g_{L} e^{-i\chi\bar
\tau_z/2},~\bar g_{R}=g_{R}, \nonumber \\ 
g_{L,R}&=&\left(\begin{array}{cc} 1-2f_{L,R}(\eps) & -2f_{L,R}(\eps)
\\ -2(1-f_{L,R}(\eps)) & 2f_{L,R}(\eps)-1 \end{array}\right),
\label{gf1}
\end{eqnarray}
where $\bar\tau_z$ is the Pauli matrix, $f_{L,R}(\eps)$ are the Fermi
distribution function of the reservoirs and $\chi$ is the counting
field (due to current conservation, it is only necessary to count the
electrons in one reservoir). The matrix $\bar g(\chi)$ is determined
from a matrix current conservation equation
\begin{eqnarray}
\bar I_{L}+\bar I_{R}=0,~\bar I_{L,R}&=&\frac{G_{L,R}[\bar
g_{L,R},\bar g]}
{4+\Gamma_{L,R}[\{\bar g_L,\bar g\}-2]}
\label{kirchoff}
\end{eqnarray}
where $[..,..]$ is the commutator and $\{..,..\}$ the
anti-commutator. Knowing $g(\chi)$, the full counting statistics of
charge transfer can be found. However, in the system under study, it
is not possible to find an explicit expression for $g(\chi)$ in the
general case (arbitrary $N_L,N_R$ and $\Gamma_L,\Gamma_R$), and the
full counting statistics has to be studied by numerical means. Here we
are only interested in the first three cumulants, which can be found
analytically by an expansion of the Green's functions in the counting
field $\chi$.

The first three cumulants are given by (evaluated at the left contact)
\begin{eqnarray}
I&=&\frac{e}{h}\int~dE~\mbox{tr}\left.[\bar \tau_z \bar
I_L]\right|_{\chi=0} \nonumber \\ \langle\langle
I_L^2\rangle\rangle&=&\left.-i\frac{e^2}{h}\int~dE~\mbox{tr}\left[\bar
\tau_z \frac{d\bar I_L}{d\chi}\right] \right|_{\chi=0} \nonumber \\
\langle\langle
I_L^3\rangle\rangle&=&-\left.\frac{e^3}{h}\int~dE~\mbox{tr}\left[\bar
\tau_z \frac{d^2\bar I_L}{d\chi^2}\right] \right|_{\chi=0}
\label{cumeq}
\end{eqnarray}
To calculate these cumulants, we thus need to expand the Green's
functions, and correspondingly the matrix currents, to second order in
the counting field $\chi$, i.e.
\begin{eqnarray}
\bar g(\chi)&=&\bar g^{(0)}+\chi \bar g^{(1)}+\frac{\chi^2}{2} 
\bar g^{(2)},~\bar g^{(n)}\equiv \left .
\frac{d^n \bar g}{d\chi^n}\right|_{\chi=0}
\end{eqnarray}
and similarly for the other quantities. For simplicity, we consider
the case with zero temperature and the left reservoir held at a finite
voltage $eV$. In this case, only the energies $0<\eps<eV$ need to be
considered, where $f_L(\eps)=1$ and $f_R(\eps)=0$, and we drop the
energy notation below.

For the first cumulant, we have the matrix currents to zeroth order in
the counting field,
\begin{eqnarray}
\bar I_{L,R}^{(0)}&=&\frac{G_{L,R}[\bar g_{L,R}^{(0)},\bar g^{(0)}]}
{4+\Gamma_{L,R}[\{\bar g_{L,R}^{(0)},\bar g^{(0)}\}-2]}.
\label{matcurr0}
\end{eqnarray}
where from Eq. (\ref{gf1}), one has $\bar g_{R,L}^{(0)}=g_{R,L}$. From
the matrix current equation in Eq. (\ref{kirchoff}),
i.e. $I_{R}^{(0)}+I_{L}^{(0)}=0$, we then obtain
\begin{eqnarray}
\bar g^{(0)}&=&\left(\begin{array}{cc} 1-2f & -2f \\ 
-2(1-f) & 2f-1 \end{array}\right),~f=\frac{G_L}{G_L+G_R}
\end{eqnarray}
where $f$ is the distribution function in the dot, as in
Eq. (\ref{f_average}). Knowing $\bar g^{(0)}$ we find $\bar I_L^{(0)}$
from Eq. (\ref{matcurr0}) and then the current from Eq. (\ref{cumeq})

For the second cumulant, we need to expand the matrix currents to
first order in $\chi$. Noting that the expressions in the matrix
denominator appearing in the expansion, is $4+\Gamma_{L,R}[\{\bar
g_{L,R}^{(0)},\bar g^{(0)}\}-2]=4$, we obtain
\begin{eqnarray}
\bar I_{L}^{(1)}&=&\frac{G_L}{4}\left([\bar g_L^{(1)},\bar
g^{(0)}]+[\bar g_L^{(0)},\bar g^{(1)}]\right)\nonumber \\ 
&+& \frac{G_L}{16}\left(\{\bar g_L^{(1)},\bar g^{(0)}\}+\{\bar
g_L^{(0)},\bar g^{(1)}\}\right),\nonumber \\ 
\bar
I_{R}^{(1)}&=&\frac{G_R}{4}[\bar g_R^{(0)},\bar g^{(1)}]+\frac{G_R}{16}\{\bar
g_R^{(0)},\bar g^{(1)}\}.
\label{matcurr1}
\end{eqnarray}
In addition to the matrix current equation $\bar I_L^{(1)}+\bar
I_R^{(1)}=0$, we get an extra condition for $\bar g^{(1)}$ from the
normalization condition $\bar g(\chi)^2=1$, namely
\begin{eqnarray}
\bar g^2(\chi)&=&1+\chi \{\bar g^{(0)}, \bar g^{(1)}\}+O(\chi^2)=1 
\nonumber \\
&\Rightarrow& \{\bar g^{(0)}, \bar g^{(1)}\}=0
\label{normcond}
\end{eqnarray}
Staring from the ansatz (equivalent to the parametrization in 
Ref. \onlinecite{Nazarov-02})
\begin{eqnarray}
\bar g^{(1)}=\left(\begin{array}{cc} h_{11}^{(1)} & h_{12}^{(1)} \\
h_{21}^{(1)} & -h_{11}^{(1)} \end{array}\right),
\end{eqnarray}
Eq. (\ref{normcond}) gives
$h_{21}^{(1)}=h_{11}^{(1)}(1-2f)/f-h_{12}^{(1)}(1-f)/f$. Inserting
$\bar g^{(1)}$ into the matrix current equation $\bar I_{L}^{(1)}+\bar
I_{R}^{(1)}=0$ gives $h_{11}^{(1)}=h_{12}^{(1)}+4f^2$ and then
\begin{eqnarray}
h_{12}^{(1)}&=&-\frac{4G_L}{(G_L+G_R)^4}\left[G_L^3+G_L^2G_R(1+\Gamma_R)
\right.
\nonumber \\ &+& \left. G_LG_R^2+G_R^3(1-\Gamma_L)\right].
\end{eqnarray}
From $h_{12}^{(1)}$ we thus obtain all components of $\bar
g^{(1)}$. Inserting this into the matrix currents in
Eq. (\ref{matcurr1}) we get the second cumulant from
Eq. (\ref{cumeq}). It coincides exactly with Eq. (\ref{I2-0}).

The calculation of the third cumulant proceeds along the same lines. One
first expands the matrix currents to second order in $\chi$ (not
presented due to the lengthy expressions). The requirement that the
$O(\chi^2)$ term in Eq. (\ref{normcond}) should be
zero gives $\{\bar g^{(2)},\bar g^{(0)}\}+2(\bar g^{(1)})^2=0$. Using
the ansatz
\begin{eqnarray}
\bar g^{(2)}=\left(\begin{array}{cc} h_{11}^{(2)} & h_{12}^{(2)} \\ 
h_{21}^{(2)} & -h_{11}^{(2)} \end{array}\right),
\end{eqnarray}
one gets
$h_{21}^{(2)}=\left[(g_a^{(1)})^2+g_b^{(1)}g_c^{(1)}\right]/f
+h_{11}^{(2)}(1-2f)/f-h_{12}^{(2)}(1-f)/f$. The
matrix current equation $\bar I_L^{(2)}+\bar I_R^{(2)}=0$ then gives
$h_{11}^{(2)}$ and $h_{12}^{(2)}$ (not written out), which fully
determines $\bar g^{(2)}$. Inserting this into the expression for the
matrix currents we find the third cumulant from Eq. (\ref{cumeq}). It
coincides exactly with Eq. (\ref{I3-0}).

We point out that it is in principle possible to obtain analytical
expressions for all higher cumulants in the same way, although the
procedure is rather cumbersome already for the third cumulant.

The third and fourth cumulants of current may be also obtained by
means of random-matrix theory\cite{Pilgram-02} using the diagrammatic
technique proposed by Brouwer and Beenakker.\cite{Brouwer-96}
Substituting the resulting transmission probabilities for the whole
system and using Eq. (\ref{chi}), one obtains expressions that
coincide with Eq. (\ref{I3-0}) and the expression for the fourth
cumulant given in the Appendix.

\section{
                     Conclusion
}\label{Summary}

In summary, we have shown that the semiclassical cascade approach
gives the same results for the third and fourth cumulants of current
in a chaotic cavity with imperfect leads as the circuit theory. This
leads us to the conclusion that this approach may be applied to a wide
class of systems that may include both semiclassical and
quantum-mechanical elements. The advantage of the cascade approach is
its physical transparency and a relative simplicity. For example, if
the system consists of a number of cavities connected by contacts
whose cumulants of current are known, this approach allows one to
easily construct the cumulants of the current for the whole system. In
principle, it also allows an inclusion of inelastic scattering
processes and a calculation of cross-correlated cumulants of current
in multiterminal systems. Therefore it presents a reasonable
alternative to the full counting statistics based on the circuit
theory.

We are grateful to M. B\"uttiker and E. V. Sukhorukov for fruitful
discussions. This work was supported by the Swiss National Foundation,
the program for Materials and Novel Electronics Properties,
Russian Foundation for Basic Research, grant 01-02-17220, and by
the INTAS Open grant 2001-1B-14. One of us is thankful to the Geneva
University for hospitality.

\vspace{1cm}
\centerline{\bf
                      APPENDIX
}

\vspace{1cm}

In the case of arbitrary transmissions of the contacts the fourth
cumulant is given by the expression
$$
 \la\la
  I_L^4
 \ra\ra
 =
 -e^3 I 
 \Bigl[
  (\Gamma_R - 1) 
  (
   6\Gamma_R^2 - 6\Gamma_R + 1
  )
  G_L^9
$$ $$
  +
  (
   60\Gamma_R^2 - 30\Gamma_R - 36\Gamma_R^3 + 5
  )
  G_R G_L^8
$$ $$
  +
  (
   30\Gamma_R^3 - 10 - 60\Gamma_R^2 + 45\Gamma_R
  )
  G_R^2 G_L^7
$$ $$
  +
  (
   120\Gamma_L\Gamma_R^2 - 60\Gamma_R^2 - 30 + 92\Gamma_R
   + 
   55\Gamma_L - 168\Gamma_L \Gamma_R
  ) 
  G_R^3 G_L^6 
$$ $$
  +
  (
   72\Gamma_L\Gamma_R + 4 + 72\Gamma_R^2 - 6\Gamma_L
   -
   96\Gamma_L\Gamma_R^2 - 51\Gamma_R
  ) 
  G_R^4 G_L^5
$$ $$
  +
  (
   4 - 6\Gamma_R - 51\Gamma_L + 72\Gamma_L^2 + 72\Gamma_L\Gamma_R
   -
   96\Gamma_L^2\Gamma_R
  ) 
  G_R^5 G_L^4 
$$ $$
  +
  (
   92\Gamma_L - 60\Gamma_L^2 - 30 - 168\Gamma_L\Gamma_R
   +
   120\Gamma_L^2\Gamma_R + 55\Gamma_R
  ) 
  G_R^6 G_L^3
$$ $$
  +
  (
   45\Gamma_L + 30\Gamma_L^3 - 10 - 60\Gamma_L^2
  ) 
  G_R^7 G_L^2
$$ $$
  +
  (
   -36\Gamma_L^3 - 30\Gamma_L + 5 + 60\Gamma_L^2
  ) 
  G_R^8 G_L 
$$ $$
  +
  (
   -1 + \Gamma_L
  )(
   6\Gamma_L^2 - 6\Gamma_L + 1
  )
  G_R^9
 \Bigr]
 \Bigl/
  (G_L + G_R)^9.
$$

\end{multicols}
\end{document}